\documentclass[12pt]{article}

\normalbaselines \topmargin -0.25 truein \oddsidemargin 0.30
truein \evensidemargin 0.30 truein 

\begin{document}

\newcommand{\nablab}{{\mathop {\rule{0pt}{0pt}{\nabla}}\limits^{\bot}}\rule{0pt}{0pt}}

\centerline{\bf  EXTENDED AXION ELECTRODYNAMICS: OPTICAL ACTIVITY}

\centerline{\bf   INDUCED BY NONSTATIONARY DARK MATTER}

 \vskip 0.3cm \centerline{\bf A.B. Balakin\footnote{e-mail: Alexander.Balakin@ksu.ru} and N.O. Tarasova\footnote{e-mail: anti-neitrino@yandex.ru}} \vskip 0.3cm
\centerline{\it Department of General Relativity and Gravitation, Kazan Federal University,}

\centerline{\it  Kremlevskaya str., 18, 420008, Kazan,  Russia}

\vskip 0.5cm
\noindent
{\bf Abstract}

\noindent {\small We establish a new self-consistent
Einstein-Maxwell-axion model based on the Lagrangian, which is
linear in the pseudoscalar (axion) field and its four-gradient
and includes the four-vector of macroscopic velocity of the axion
system as a whole. We consider extended equations of the axion
electrodynamics, modified gravity field equations, and
discuss nonstationary effects in the phenomenon of optical
activity induced by axions.}

\section{Introduction}

Axions (pseudo-Goldstone bosons) are considered to be Weakly
Interacting Massive Particles (WIMPs) appearing as a result of
spontaneous phase transition predicted by Peccei and Quinn
\cite{Peccei}. These (hypothetic) particles can play a fundamental
role in the formation of Dark Matter, whose contribution into the
Universe energy balance is estimated to be about 23 \% (see, e.g.,
\cite{Raffelt} - \cite{Battesti}). The axion electrodynamics
established by Weinberg and Wilczek \cite{Weinberg}-\cite{Wilczek}
gives us a new instrument for the Dark Matter investigation, since
the model of coupling between photons and pseudoscalar (axion)
field proposed by Ni in  \cite{8Ni} predicts the effect of
polarization rotation, when the electromagnetic waves travel
through the axion system (see, e.g., \cite{19Ni}-\cite{24}). The
optical activity is the best known but not unique phenomenon,
which can be induced by axions in the electrodynamic systems. We
expect that birefringence, dynamo-optical, etc. effects, which are
well-known in the classical electrodynamics of moving media, can
be found in axion-photons systems as well. Keeping in mind this
idea, we intend to establish a number of new models unified
terminologically by the common title "Extended Axion
Electrodynamics". Non-minimal extension of the
Einstein-Maxwell-axion theory \cite{BaWTN} was the first step in
that direction. In this short note we discuss the class of models,
which satisfy the following requirements: first, the
electrodynamics is linear; second, the pseudoscalar field equation
is of the second order in derivatives; third, the cross-terms in
the Lagrangian are linear in the pseudoscalar field and its
four-gradient; fourth, the Lagrangian of the model includes the
macroscopic velocity four-vector of the system as a whole, but
does not contain its derivatives.

\section{Extended model of the axion-photon coupling}

\subsection{On the Lagrangian of the extended model}

The {\it standard} Einstein-Maxwell-axion model is based on the Lagrangian formalism with the action functional
\begin{equation}
S_{(0)} = \int d^4 x \sqrt{{-}g} \ {\cal L}_0 \,,
\label{action01}
\end{equation}
\begin{equation}
{\cal L}_{(0)} = \frac {R{+}2\Lambda}{\kappa}{+}\frac{1}{2}F_{mn}F^{mn} {+}\frac{1}{2}\phi F^{*}_{mn} F^{mn}
{+} \Psi_{0}^2\left[ {-}\nabla_m\phi \nabla^m\phi {+}V(\phi^2) \right]
\,.
\label{action02}
\end{equation}
Here, $g$ is the determinant of the metric tensor $g_{ik}$,
$\nabla_{m}$ is the covariant derivative, $R$ is the Ricci scalar, $\kappa \equiv \frac{8\pi G}{c^4}$ is the Einstein coupling constant,
$\Lambda$ is the cosmological constant.
The Maxwell tensor $F_{mn}$ is given by
\begin{equation}
F_{mn} \equiv \nabla_m A_{n} - \nabla_n A_{m} \,, \quad \nabla_{k}
F^{*ik} =0 \,, \label{maxtensor}
\end{equation}
where $A_m$ is an electromagnetic potential four-vector; $F^{*mn}
\equiv \frac{1}{2} \epsilon^{mnpq}F_{pq}$ is the tensor dual to
$F_{pq}$; $\epsilon^{mnpq} \equiv \frac{1}{\sqrt{-g}} E^{mnpq}$ is
the Levi-Civita tensor, $E^{mnpq}$ is the absolutely antisymmetric
Levi-Civita symbol with $E^{0123}=1$. It is the third term in the
Lagrangian that describes the pseudoscalar-photon interaction
\cite{8Ni}. The symbol $\phi$ stands for a pseudoscalar field,
this quantity being dimensionless. The axion field itself, $\Phi$, is
considered to be proportional to this quantity $\Phi = \Psi_0
\phi$ with a phenomenological constant $\Psi_0$. The function
$V(\phi^2)$ describes the potential of the pseudoscalar field.

Now we extend the Lagrangian (\ref{action02}) by the terms, which are quadratic in the Maxwell tensor $F_{mn}$, are linear in
$\phi$ or in $\nabla_k \phi$, and contain the normalized four-vector $U^k$ $(U^k U_k=1)$.
The quantity $U^k$ describes the macroscopic velocity of the axion system as a whole, and may be chosen as the time-like
eigen-vector of the stress-energy tensor of the pseudoscalar (axion) field. In order to list all the irreducible scalars,
which satisfy these requirements, let us remind the important identity
\begin{equation}
F^{ik} F^{*}_{kj} = \frac{1}{4}\delta^i_j \ F^{mn} F^{*}_{mn}  \,. \label{star0}
\end{equation}
Clearly, all the invariants, which we could construct using $g_{ij}$, $F^{ik}$, $F^{*}_{mn}$, $U^k$, as well as, $\phi$ or $\nabla_k \phi$,
definitely contain at least one convolution of the type $F^{ik} F^{*}_{kj}$. Thus, it is easy to check that due to (\ref{star0}) all the
new terms in the extended Lagrangian can be reduced to the invariant
\begin{equation}
{\cal L}_{({\rm int})} = \frac{1}{2} \nu \ F^{mn} F^{*}_{mn} \ U^k \nabla_k \phi  \,, \label{star3}
\end{equation}
where $\nu$ is some new coupling constant introduced phenomenologically.

\subsection{Extension of the axion electrodynamics}

The variation of the action functional containing the sum of Lagrangians  ${\cal L}_{(0)}{+}{\cal L}_{({\rm int})}$ with respect
to the four-vector potential $A_i$ gives the equations of axion electrodynamics
\begin{equation}
\nabla_k H^{ik}=0 \,.
\label{eld1}
\end{equation}
Here the excitation tensor $H^{ik}$ is given by the term
\begin{equation}
H^{ik}= F^{ik} {+}  F^{*ik} \left(\phi + \nu {\cal D} \phi \right)  \,,
\label{eld2}
\end{equation}
and ${\cal D}=U^k \nabla_k$ is the convective derivative.
Using the linear constitutive equations
\begin{equation}
H^{ik} = C^{ikmn} F_{mn} \,,
\label{eld3}
\end{equation}
we readily obtain that the linear response tensor $C^{ikmn}$ now takes the form
\begin{equation}
C^{ikmn} = \frac{1}{2} \left( g^{im} g^{kn} {-} g^{in} g^{km} \right) {+} \frac{1}{2} \epsilon^{ikmn} \left(\phi + \nu {\cal D} \phi \right)
\,.
\label{eld4}
\end{equation}
This means that the dielectric permittivity and magnetic impermeability tensors of the axion-photon system are the same as in vacuum, i.e.,
\begin{equation}
\varepsilon^{im} = 2 C^{ikmn} U_k U_n = \Delta^{im} \,,
\label{eld5}
\end{equation}
where $\Delta^{im}=g^{im}-U^i U^m$ is the projector, and
\begin{equation}
(\mu^{-1})_{pq}  = - \frac{1}{2} \eta_{pik} C^{ikmn}
\eta_{mnq} = \Delta_{pq} \,,
\label{eld6}
\end{equation}
where $\eta_{pik} \equiv \epsilon_{pikj}U^j$.
The tensor of magneto-electric coefficients
\begin{equation}
\nu_{p}^{\ m} = \eta_{pik} C^{ikmn} U_n = - \Delta_p^m \left(\phi + \nu {\cal D} \phi \right) \,, \label{eld7}
\end{equation}
describing optical activity effects (see, e.g., \cite{nu1}) is now characterized by additional term linear in ${\cal D}\phi$.

\subsection{Pseudoscalar field evolution}

Variation of the extended action functional with respect to the pseudoscalar field $\phi$ gives the equation
\begin{equation}
\nabla_k \nabla^k \phi + \phi V'(\phi^2) =
\frac{1}{4\Psi_{0}^2}\left[ F^{*}_{mn}F^{mn} \left(\nu \theta
-1\right) + \nu {\cal D} \left(F^{*}_{mn}F^{mn}\right) \right]\,,
\end{equation}
where $\theta \equiv \nabla_k U^k$ is the expansion scalar of the
velocity field, and the prime denotes the derivative of the potential $V(\phi^2)$ with respect
to the argument.

\subsection{Gravity field equations}

Modified Einstein's equations obtained by the variation of the extended action functional with respect to the metric $g^{pq}$
can be presented in the form
\begin{equation}
R_{pq} {-} \frac{1}{2} g_{pq} R =  \Lambda g_{pq} + \kappa \left[
T_{pq}^{(EM)} + T_{pq}^{(A)} + \nu T_{pq}^{(*)} \right]\,.
\label{gr1}
\end{equation}
Here the stress-energy tensor of the electromagnetic field
\begin{equation}
T_{pq}^{(EM)} = \frac{1}{4} g_{pq} F_{mn} F^{mn} - F_{pm} F_{q}^{\ m}
\label{gr2}
\end{equation}
and the stress-energy tensor of the pure axionic field
\begin{equation}
T_{pq}^{(A)} =\Psi_{0}^2 \left\{ \nabla_p \phi \nabla_q \phi -  \frac{1}{2} g_{pq} \left[ \nabla_m \phi \nabla^m \phi
- V\left( \phi^2 \right) \right] \right\}
\label{gr3}
\end{equation}
are presented by the well-known terms. The tensor
\begin{equation}
T_{pq}^{(*)} =  - \frac{1}{8} F^{mn}F^{*}_{mn} \left(U_p\nabla_q \phi {+}U_q \nabla_p \phi \right)  \,,
\label{gr4}
\end{equation}
describes a principally new source-term in the right-hand side of the gravity field equations.
Let us mention that the term $\nu T_{pq}^{(*)}$ is obtained by the variation of the term with the interaction Lagrangian (\ref{star3})
 by using the formula
\begin{equation}
\delta U^i =  \frac{1}{4} \delta g^{pq} \left(U_p \delta^i_q  {+}
U_q \delta^i_p \right) \label{gr5}
\end{equation}
for the variation of the velocity four-vector (see \cite{GraCos} for details). The standard interaction term
$\frac{1}{2}\sqrt{{-}g} \phi F^{mn}F^{*}_{mn} = \frac{1}{4} \phi E^{ikmn}F_{ik}F_{mn}$ does not contribute to the stress-energy tensor
in the process of variation with respect to the metric. Thus, the appearance of the term (\ref{gr4}) is a new event in the modeling of the
gravity field of the photon-axion system.

\subsection{An example of application}

Let us consider the propagation of {\it test} electromagnetic wave
coupled to the axionic subsystem of the Dark Matter in the spatially homogeneous
FLRW-type spacetime with the scale factor $a(t)$. Let the
electromagnetic wave propagate in the direction $0x$ and be
characterized by the potential four-vector
$A_{i}{=} \delta_i^2 A_2(t,x){+}\delta_i^3 A_3(t,x)$.
The equations of axion electrodynamics can be now reduced to
\begin{equation}
\left[\frac{\partial^2}{\partial t^2} -
\frac{1}{a^2}\frac{\partial^2}{\partial x^2} + H
\frac{\partial}{\partial t} \right] A_2 = -
\frac{2\dot{\Theta}}{a} \frac{\partial}{\partial x} A_3 \,,
\label{S1}
\end{equation}
\begin{equation}
\left[\frac{\partial^2}{\partial t^2} -
\frac{1}{a^2}\frac{\partial^2}{\partial x^2} + H
\frac{\partial}{\partial t} \right] A_3 = \frac{2\dot{\Theta}}{a}
\frac{\partial}{\partial x} A_2 \,,\label{S11}
\end{equation}
where $H(t)=\frac{\dot{a}}{a}$ is the Hubble function and
\begin{equation}
\Theta(t) \equiv \frac{1}{2}\left[\phi(t) + \nu \dot{\phi}(t)\right]
\,,\label{S111}
\end{equation}
(in the cosmological context we use the units with $c=1$).
Clearly, when $\nu =0$ and $\dot{\phi}\neq 0$ the electromagnetic wave can not keep linear polarization, however,
in case when $\nu \neq 0$ and the pseudoscalar field evolves exponentially $\phi(t) \propto \exp{\left[-\frac{t}{\nu}\right]}$, it can be possible.
In the approximation of short wavelengths $k>>H$ the solution of (\ref{S1}), (\ref{S11}) for the circularly polarized wave has the form
\begin{equation}
A_2 = - A_0 \sin{\left[W - \varphi(t)\right]} \,, \quad A_3 = A_0
\cos{\left[W - \varphi(t) \right]} \,,\label{S6}
\end{equation}
\begin{equation}
 W = W(t_0) + k\left[\int_{t_0}^t \frac{dt'}{a(t')} - x \right] \,, \quad \varphi(t) \equiv \Theta(t)-\Theta(t_0)
\,,\label{S21}
\end{equation}
where $k$ is a constant reciprocal to the wavelength. On the one
hand, the quantity $\varphi(t)$ is expressed in terms of
$\phi$ and $\dot{\phi}$ (see (\ref{S21}) and (\ref{S111})) and
describes the rotation angle of the polarization vector of the
electromagnetic wave traveling through the axion system; it can be
studied in optical experiments. On the other hand, it is well-known
that in the cosmological context the function $\dot{\phi}$ can be
represented in terms of the Dark Matter energy-density ${\cal E}$
and pressure ${\cal P}$ as follows
\begin{equation}
\dot{\phi} = \pm \frac{1}{\Psi_0} \sqrt{{\cal E}(t)+{\cal P}(t)} \,.
\label{cl13}
\end{equation}
Thus, the extended axion electrodynamics can be considered as a
tool for investigation of the nonstationary effects in the
evolution of the axionic Dark Matter.

\vspace{5mm}
\noindent {\large\bf Acknowledgments} The authors are
grateful to Dr. Zayats A.E. for fruitful comments. This work was
supported by the FTP "Scientific and Scientific-Pedagogical
Personnel of the Innovative Russia" (grants Nos 16.740.11.0185 and
14.740.11.0407), and by the RFBR (grants Nos. 11-02-01162 and 11-05-97518-p-center-a).

\end{document}